\documentclass[preprint,showpacs,aps,prb,amsmath,amssymb]{revtex4}
\usepackage{amsmath}
\usepackage{graphicx}
\begin{document}

\title{Magnetism in one-dimensional quantum dot arrays}

\author{K. K\"arkk\"ainen, M. Koskinen, S.M. Reimann$^*$ and M. Manninen}
\affiliation{Nanoscience Center, Department of Physics,
FIN-40014 University of Jyv\"askyl\"a, Finland}
\affiliation{$^*$Mathematical Physics, 
Lund Institute of Technology, SE-22100 Lund, Sweden}

\date{\today}

\begin{abstract}
We employ the density functional Kohn-Sham method in the local 
spin-density approximation to study the electronic structure and magnetism 
of quasi one-dimensional periodic arrays of few-electron quantum dots.
At small values of the lattice constant, the single dots overlap, 
forming a non-magnetic quantum wire with nearly homogenous density.
As the confinement perpendicular to the wire is increased, i.e. as 
the wire is squeezed to become more one-dimensional, 
it undergoes a spin-Peierls transition.  
Magnetism sets in as the quantum dots are placed further apart. It 
is determined by the electronic shell filling of the individual quantum 
dots. At larger values of the 
lattice constant, the band structure for odd numbers of electrons 
per dot indicates that the array could support spin-polarized transport
and therefore act as a spin filter.
\end{abstract}

\maketitle

\section{Introduction}

Quantum dots or ``artificial atoms'', as they are frequently called,  
confine a few electrons on a small conduction electron island, 
built in (or from) a semiconductor heterostructure. 
Being finite-sized fermion systems, quantum dots can show strong 
shell effects which determine their physical properties. 
Just like for atoms, quantum dots with closed shells 
are particularly stable, implying ``noble'' structures for certain numbers 
of electrons in the dot.  Following Hund's rules, at half-filling of a shell, 
orbital degeneracy can lead to spin alignment.
This was discovered first for small vertical quantum dot samples 
with circular-parabolic shape by Tarucha {\it et al.}~\cite{tarucha96}.
The experimental findings were later theoretically 
confirmed by electronic structure calculations using mean field methods as
well as quantum Monte Carlo techniques or even a numerical diagonalization
of the full many-body Hamiltonian (see Reimann and Manninen~\cite{reimann02} 
for a review). 

Experimentally, few-electron quantum dot structures where the shell
effects on magnetism could be observed, are challenging 
to fabricate. One example where spontaneous magnetism has been found, 
are one-dimensional quantum point contact constrictions formed 
in a gate-patterned heterostructure \cite{thomas96,pyshkin00,heyman}. 
The intrinsic magnetic properties of these nanostructures have 
drawn much attention recently 
due to their potential applicability in spintronics devices \cite{zutic04}. 
Quantum point contacts \cite{potok02} 
and single quantum dots \cite{folk03,fransson04} were found to 
have spin filtering capabilities, with a possibility to  
serve for either generating or detecting spin-polarized currents.

Arranging many quantum dots in a lattice, one can 
build artificial crystals with designed band 
structure~\cite{kowenhoven1990,haug} which 
can be manipulated for example by tuning the inter-dot coupling 
and the number of confined electrons in the single quantum dots. 
The dot lattice does not suffer from structural deformations, 
which has the advantage that it can be designed freely without 
having to consider lattice instabilities~\cite{tamura2002}.

Fabrication of a quasi one-dimensional artifical crystal 
consisting of a sequence of a few quantum dots was suggested by 
Kouwenhoven {\it et al.}~\cite{kowenhoven1990} already in 1990. 
They observed oscillations in the conductance as a function of 
gate voltage, arising from the mini-band structure in the 
periodic crystal. 
Small dots in well-ordered lattices could be synthesized by 
self-organized growth~\cite{temmyo}. A particularly interesting 
artificial lattice structure is the Kagome lattice, 
due to the possibility of flat-band 
ferromagnetism~\cite{shiraishi,kimura,kimura2002,tamura2002,mohan}.
Shiraishi {\it et al.}~\cite{shiraishi} 
have pointed at the importance of these structures 
for fast processing and high-density storage of information.

For square lattices, Koskinen \emph{et al.}\cite{koskinen03} 
showed within the density-functional
scheme that few-electron quantum dot lattices  
have a rich magnetic phase diagram, depending on the 
lattice constant and electron number.
Related observations have been made also within the 
Hubbard model~\cite{chen1997,tamura2002,koskinen2003}.

In this paper, we  investigate the electronic and magnetic properties of 
quasi one-dimensional quantum dot arrays. We suggest that such 
linear quantum dot chains could, in fact, lead to single-spin 
conductivity. 

In our model the single quantum dot 
confinement is provided by a rigid Gaussian-shaped 
background charge distribution.  At the single dot centers, this 
potential is approximately parabolic. 
The band structure and the magnetic properties depend on the lattice 
constant, $a$, and the number of electrons per dot,
$N$. Here, conductivity of the dot chain is only considered by observing 
whether there is a band gap at the Fermi-level or not, which allows a 
qualitative understanding. 

At small values of the lattice constant, the single dots overlap, 
forming a non-magnetic quantum wire with nearly homogenous density.
As the confinement perpendicular to the wire is increased, i.e. 
the wire is squeezed to become more one-dimensional, 
the ground state is a spin density wave caused by a spin-Peierls transition\cite{peierls1955,hook1991,reimann1999}. 
Magnetism sets on as the lattice constant is increased.
It is determined by the shell structure of the individual dots: 
the arrays are non-magnetic insulators 
for closed single-dot shells at $N=2$ and 6. 
At half-filled shell ($N=4$) the spin of the dot is determined by
Hund's rule and the array is an antiferromagnetic insulator. Ferromagnetism is
observed both at the beginning and the end of a shell
(here $N=3,5$). The spin-up and spin-down bands are separated by the
exchange splitting. At sufficiently large lattice constant 
$a$ one observes a gap between these bands. 
In this case the current would be carried by a single spin only, 
acting as a spin filter. 

\section{The computational method}

In order to model the one-dimensional quantum dot array, we consider interacting
electrons moving in two dimensions in a rigid periodic
background charge distribution $e\rho_B$.
The background charge number per unit cell is chosen 
to match the electronic charge of the unit cell in order to ensure overall charge neutrality. 
We employ the Kohn-Sham method with periodic boundary conditions.
The Kohn-Sham orbitals are of Bloch form 
$\psi_{n\mathbf{k}\sigma}(\mathbf{r})=\exp(i\mathbf{k}\cdot\mathbf{r})u_{n\mathbf{k}\sigma}(\mathbf{r})$, 
where $n$ labels the band, $\sigma=(\downarrow,\uparrow)$ is the spin index 
and the wave vector $\mathbf{k}$ is confined into the first
Brillouin zone. The periodic functions $u_{n\mathbf{k}\sigma}(\mathbf{r})$ satisfy the Bloch-Kohn-Sham equations
\begin{equation}
-\frac{\hbar^2}{2m^*}(\nabla+i\mathbf{k})^2u_{n\mathbf{k}\sigma}
(\mathbf{r})+v_{eff}^{\sigma}(\mathbf{r})u_{n\mathbf{k}\sigma}(\mathbf{r})
=\varepsilon_{n\mathbf{k}\sigma}u_{n\mathbf{k}\sigma}(\mathbf{r})
\end{equation}
where the periodic effective potential is
\begin{equation}
v_{eff}^{\sigma}(\mathbf{r})=\int\frac{e^2(\rho(\mathbf{r}')-
\rho_B(\mathbf{r}'))}{4\pi\epsilon_0\epsilon|\mathbf{r}
-\mathbf{r}'|}d\mathbf{r}'+v_{xc}^{\sigma}[\rho(\mathbf{r}),\xi(\mathbf{r})],
\end{equation}
$\rho$ is the electron density and
$\xi=(\rho_\uparrow-\rho_\downarrow)/\rho$ is the polarization.
In the local spin-density approximation we use the generalized~\cite{sdw} 
Tanatar-Ceperley~\cite{tc} parameterization for the polarization-dependent 
exchange-correlation potential $v_{xc}^{\sigma}[\rho(\mathbf{r}),\xi(\mathbf{r})]$.
In the band structure calculation, the functions $u_{n\mathbf{k}\sigma}(\mathbf{r})$ 
are expanded in a basis with $11\times 11$ plane waves. 
For one-dimensional systems, the wave vector reduces to wave number for which we chose 
an equidistant 19-point mesh in the first Brillouin zone. 
The self-consistent iterations were started with anti-ferromagnetic and ferromagnetic initial potentials. 
Small random perturbations were added to the initial guesses in order to avoid convergence 
into saddle points of the potential surface. In addition, 
we use an artificial temperature to 
allow fractional occupation numbers for nearly degenerate states at the Fermi
level. 
We noted that by decreasing the temperature the amplitudes of the
spin-density and the average spin per dot               
become somewhat higher for small lattice constants. Nevertheless, we must emphasize that 
the temperature is low enough not to affect the ground-state.
The statistical occupations merely help occupying degenerate levels to ensure convergence. 
We use effective atomic units with Hartree $\textrm{Ha}=m^*e^4/\hbar^3(4\pi\epsilon_0\epsilon)^2$ for energy and the 
Bohr radius $a_B^*=\hbar^2 4\pi\epsilon_0\epsilon/m^*e^2$ for 
length, where $m^* $ is the effective mass and $\epsilon $ the 
dielectric constant of the semiconductor material in question. 

\section{Magnetism in a 1D quantum dot array}

Studying magnetism in a one-dimensional array, the simplest geometry  
to choose for the unit cell is a rectangle with two quantum dots 
per cell. These
dots lie in a row along the $x$ axis of the cell, one in the center 
and one crossing periodically the edge of the cell. 
The confining potential is modeled by a periodic positive 
background charge distribution described by a sum of 
Gaussians centered at lattice sites $\mathbf{R}=a(n_x,0)$, $n_x=0,1,2,\ldots$,
\begin{equation}
\rho_B(\mathbf{r})=\sum_{\mathbf{R}}\rho_d(\mathbf{r}-\mathbf{R});\quad 
\rho_d(\mathbf{r})=\frac{1}{\pi r_s^2}\exp(-r^2/Nr_s^2),
\end{equation}
where ${\bf r}=(x,y)$ is a two-dimensional position vector. 
A single Gaussian carries positive charge $Ne$ with density $1/\pi
r_s^2$ at the center.
The parameter $r_s$ determines the average electron density at the
center of the dot. Throughout this paper we use the value $r_s=2\ a_B^*$
which is close to the equilibrium density of the two-dimensional
electron gas.
The bottom of the confining potential provided by the background charge 
distribution is harmonic to a good approximation. 
Since there are two quantum dots in the unit cell, the electronic levels are 
split into bonding and anti-bonding bands. As a consequence, for both spins 
there are two 1s-bands, four p-bands, six 2s1d-bands and so on.
In a one-dimensional quantum dot array one can have a smooth transition from the 
tight-binding description to the nearly-free electron picture simply by varying the lattice constant $a$. 

\begin{figure}
\includegraphics[width=0.8\textwidth]{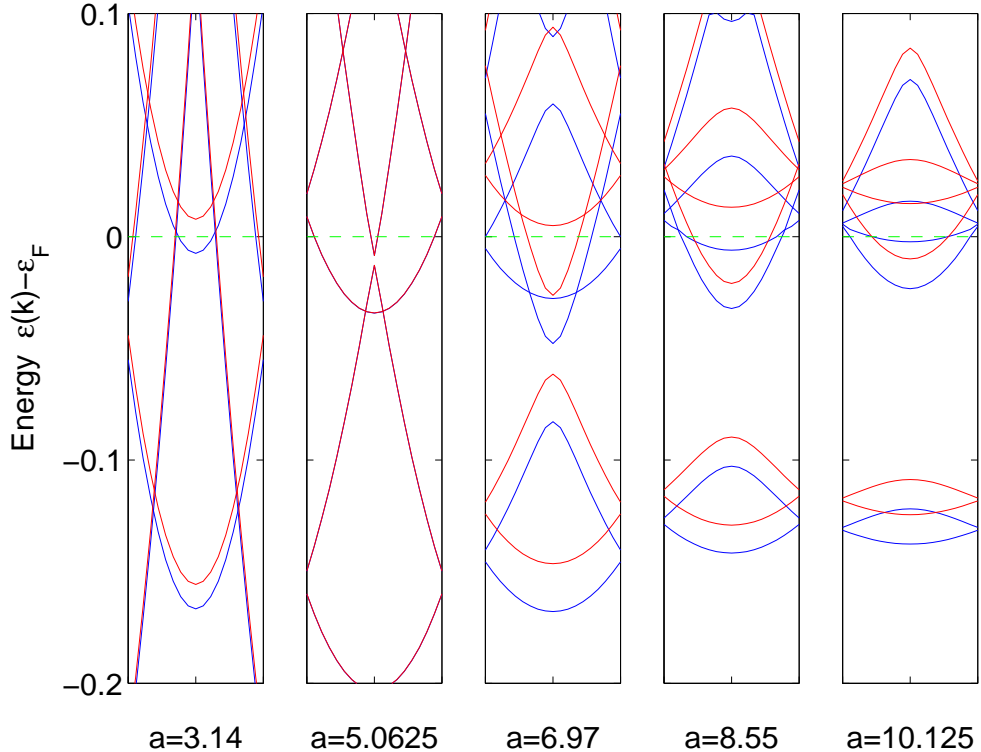}
\caption{Lowest bands at selected values of the lattice constant $a$ 
for a quantum dot array with three electrons per quantum dot
(in atomic units, see text).
The spin-down bands are plotted in blue color, and spin-up bands are plotted
in red. The dashed green line indicates the Fermi-level fixed at zero energy.}
\label{fig5}
\end{figure}

Figure \ref{fig5} shows the bands for $N=3$ with different inter-dot
separations. Spin-up and spin-down bands are plotted in red and blue color, 
respectively, and the Fermi-level is fixed at zero energy.
The spin degeneracy is lifted by the exchange splitting.
For very large values of the lattice constant $a$, 
the electron densities of the single dots hardly overlap, and the dots are 
isolated.

The energies of ferromagnetic and anti-ferromagnetic solutions are nearly degenerate as the local approximation is unable to
distinguish between them. Furthermore, the bands are flat with band gap energies 
approximately equal to the single dot level spacings. 
Even though the Fermi energy 
stays inside a band, 
the dot array becomes an insulator due to a diminished hopping
probability between the single dots in this limit. 

By decreasing $a$, i.e. by bringing the quantum dots closer to one another, 
the band dispersion increases. 
The bands corresponding to some specific quantum dot level are bunched and the 
bunches are separated by energy gaps, which is demonstrated in figure
\ref{fig5} for lattice constant $a=10.125\ a_B^*$. 
By decreasing $a$ further, 
the band gaps close. For sufficiently small $a$ the quantum dots overlap 
strongly, which leads to an essentially homogenous quantum wire with a 
Gaussian cross-section. 
In this nearly-free limit the transverse motion separates from the
longitudinal one. Consequently, the transverse states are quantized by 
the Gaussian-shaped well, while the longitudinal states remain ``free'' 
with parabolic dispersion. 
This is reflected in the band structure,  
showing nearly equidistant sub-band parabolae
where the $n$th sub-band for a given $k$ corresponds to
a Kohn-Sham-Bloch orbital with $n-1$ nodes in the transverse direction. 
We note also that the higher bands have parabolic dispersion at longer inter-dot separations than
the lower ones due to the longer spatial extent of high-energy orbitals.
>From figure \ref{fig5} we note that the second transverse sub-band is occupied at $a=5.0625\ a_B^*$
while at $a=3.14\ a_B^*$ the Fermi-level reaches the third sub-band. 
Having now also the higher transverse modes occupied, 
the quantum dot chain becomes {\it quasi} one-dimensional.

\begin{figure}
\includegraphics[width=0.8\textwidth]{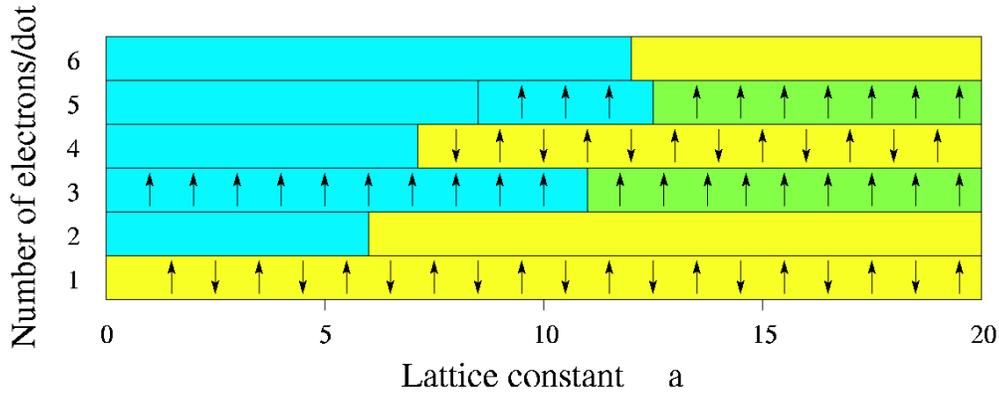}
\caption{Magnetism in a linear chain of quantum dots as a function of 
the number of electrons per dot, and the lattice constant. 
Blue color corresponds to the conducting, yellow to the insulating 
phase. Green indicates the phase where 
only one spin is conductive.}
\label{fig1}
\end{figure}

Figure \ref{fig1} shows the magnetism of the quasi-1D quantum dot array as a 
function of electron number per quantum dot and lattice constant $a$.
The colors indicate the regions where the array is conducting (blue) or insulating (yellow). 
The green bars indicate regions where the Fermi-level
resides solely on a single spin band. The arrows indicate the spin arrangement in the array.

\begin{figure}
\includegraphics[scale=0.7]{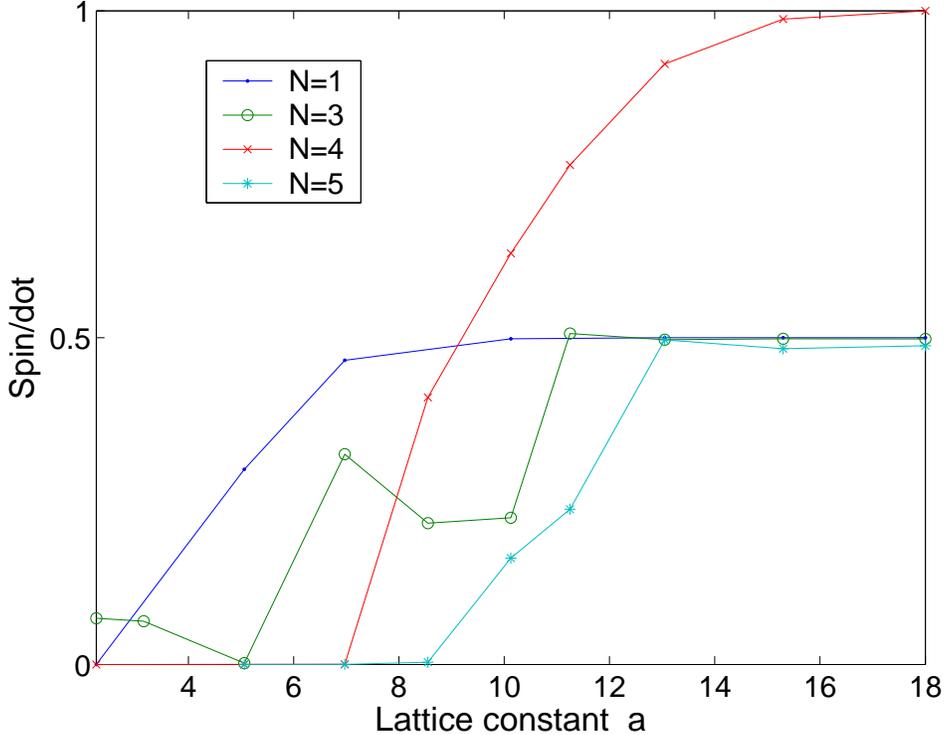}
\caption{Spin per dot for $N$=1,3,4 and 5 as a function of lattice constant.}
\label{fig2}
\end{figure}

For a single electron per quantum dot, $N=1$ only the bonding s-band is filled. 
Due to the exchange splitting of the single dot levels, the bonding and anti-bonding bands are separated by 
an energy gap and the array shows anti-ferromagnetic order. Figure \ref{fig2} shows that the 
average spin per dot, calculated by integrating the spin density over a single dot, drops gradually 
from $1/2$ to 0 as the lattice constant is decreased.
The band gap and thus the antiferromagnetism persists down to  very small
values of the lattice constant. 
At the closed shell $N=2$ the bonding and anti-bonding 1s-bands are filled
leading to a non-magnetic insulator. 
The transition from a tightly bound insulator to a nearly free metal occurs at 
the lattice constant $a\approx 6\ a_B^*$, when the gap between the 1s- and p-bands closes up.

\begin{figure}
\includegraphics[scale=0.7]{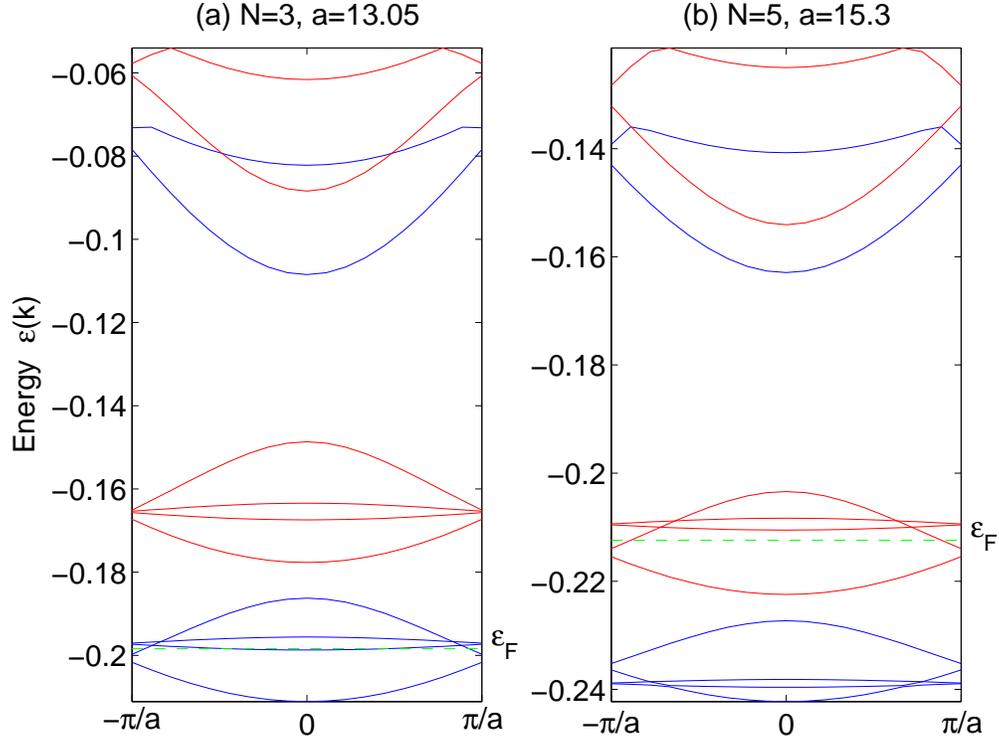}
\caption{1p-bands and lowest 2s1d-bands for (a) $N=3$ at $a=13.05\ a_B^*$ and
  (b) $N=5$ at $a=15.3\ a_B^*$. The blue lines are the 
spin-down bands, the red lines the spin-up bands. 
The dashed green line indicates the Fermi level.
The spin-up and spin-down bands are separated by a gap which leads to single-spin conductivity.}
\label{fig3}
\end{figure}

\begin{figure}
\includegraphics[width=0.8\textwidth]{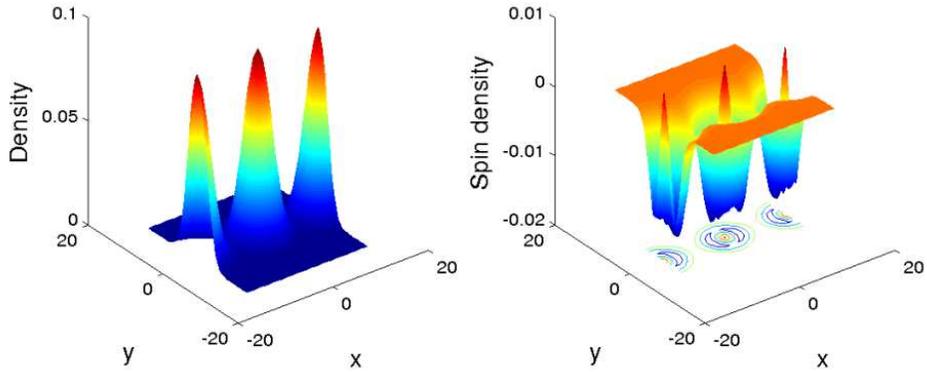}
\caption{Total density and spin density for $N=3$ at lattice constant $13.05\ a_B^*$.}
\label{fig4}
\end{figure}

Next, the p-bands are occupied. There are two degenerate p-orbitals for a single dot giving 
rise to two bonding bands and two anti-bonding bands for both the spins 
as shown in figure \ref{fig3}. The orbitals with density lobes 
oriented along the wire yield higher dispersion than the ones perpendicular to it. 
For $N=3$ there is one p-electron per quantum dot, which triggers
ferromagnetism. An example of the total electron and spin densities
is shown in Fig. \ref{fig4}.
The levels with majority spin are lower than the ones with minority 
spin as a result of the exchange splitting of the energy bands.
The density in the array increases as the dots are brought closer. 
Consequently, the 
kinetic energy becomes the dominant contribution to the total energy.
>From figure \ref{fig5} we note that at $a=5.0625\ a_B^*$ the dispersion is parabolic and the spin degeneracy is
restored. 
However, there is a competition between the 
kinetic and exchange energies. It turned out that at $a=3.14\ a_B^*$ a small spin-splitting is re-gained.

Since the Fermi-level is bound to the p-band region, the array with three electrons per dot
remains conductive at all lattice constants. 
Figure \ref{fig3} shows that the bands of the minority spin are pushed
up in energy by exchange splitting. As a consequence,
just before the insulating phase, when the bandwidths are relatively narrow, 
the Fermi-level for the minority spin lies in the band-gap but
the majority spin remains conductive. A similar behavior is found in the case of $N=5$ 
since there are three p-electrons and the shell is almost filled.
However, this time the minority spin is conductive because the levels with
fewer electrons are pushed to higher energies.  
The spin-dependent conductivity might open an intriguing opportunity to use
the linear quantum dot chain as a spin filter.

At half-filled p-shell ($N=4$) Hund's rule leads to maximized spin in an isolated quantum dot. 
Indeed, the spin per dot for $N=4$ is at its maximum (1.0) at 
lattice constant $a\approx 18\ a_B^*$ and it decreases gradually 
with $a$ as the spin densities ''spill'' into the other dots. 
Due to Hund's rule and the exchange-splitting, the anti-ferromagnetic ordering of spins is favored. 
Since a gap is formed at the Fermi-surface, the array 
is insulating until a transition to a nearly homogenous wire occurs. 
Finally for $N=6$, the p-shell is filled and the linear quantum dot chain 
remains non-magnetic at all values of $a$.

\section{Spin-Peierls transition in a homogenous quantum wire}
At small values of the 
lattice constant $a$ the quantum dots overlap significantly,  
forming a homogenous quantum wire with a Gaussian cross-section. Let us look at this limit more closely.
Consider a quantum wire with a Gaussian cross-section closed in a rectangular unit cell.
The background charge distribution is chosen to be
\begin{equation}
\rho_B(x,y)=\frac{1}{2r_s^{1D}}\frac{1}{\sqrt{2\pi}\alpha}\exp(-\frac{y^2}{2\alpha^2}),
\end{equation}
where $r_s^{1D}$ is the one-dimensional density parameter. The wire lies along 
the $x$-axis and
its width is measured by the full width at half maximum $2\sqrt{2\ln 2}\alpha$.
Since there is no definite lattice parameter for the wire, the length $L$ of the unit cell is chosen 
such that $\rho_B$ integrates to the desired charge $Ne$, thus we
have $L=2r_s^{1D}N$.  We have chosen four electrons in the unit cell ($N=4$) and fixed $r_s^{1D}=2\ a_B^*$. 
In addition, we define parameter $C_{1D}=2r_s^{1D}/\alpha$ to describe the ratio of the average 
inter-electron separation and the width of the wire: with increasing $C_{1D}$ the wire becomes narrower.
Consequently, the energies of the higher transverse modes are pushed up in energy.

\begin{figure}
\includegraphics[width=0.78\textwidth]{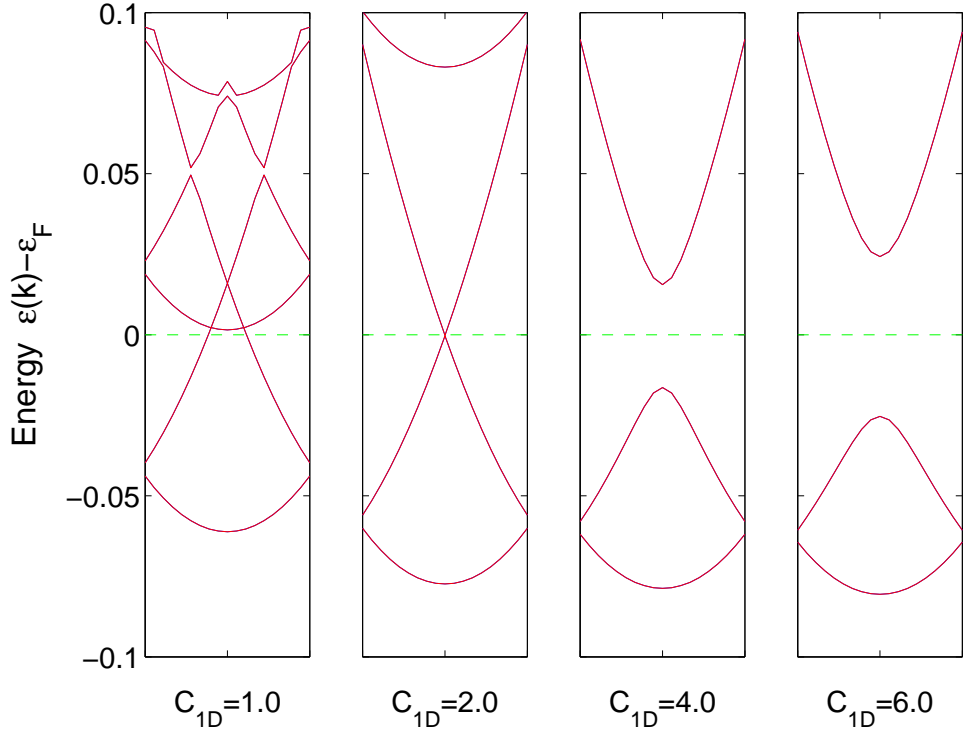}
\caption{Lowest bands at selected values of width parameter $C_{1D}$ for a quantum wire with four electrons per unit cell.
The dashed green line indicates the Fermi-level fixed at zero energy.}
\label{fig6}
\end{figure}

Figure \ref{fig6} shows band structures of homogenous quantum wires for selected widths. 
For $C_{1D}=2$, the dispersion is parabolic and the Fermi-level lies close to the 
second transverse sub-band. In this case the wire shows no magnetism.
Antiferromagnetism sets on at $C_{1D}=4$, as the spin-Peierls transition occurs. 
The ground state is a spin density wave with wave length of $L/2=r_s^{1D}N=8\ a_B^*$. 
The spin-Peierls transition opens a gap at the Fermi-level and turns the wire into an insulator. 
The amplitude of the spin density wave increases with increasing $C_{1D}$.

\section{Summary}

We studied the electronic and magnetic properties of one-dimensional arrays of few-electron quantum dots.
The spin per dot, and thus the magnetism of the array, depends on the shell filling of the individual
dots and the inter-dot coupling. Furthermore the band structure of chains with
open-shell dots
suggests that conductivity could become spin dependent at suitable values of the
lattice constant. 

\acknowledgments{We thank M. Borgh for helpful discussions. 
This work was financially supported by the Academy of Finland, 
the European Community project ULTRA-1D (NMP4-CT-2003-505457), 
the Swedish Research Council and 
the Swedish Foundation for Strategic Research.}

\end{document}